# Brazilian National High School Examination: assessing student's learning in Physics

**Marta F. Barroso**
*UFRJ – Federal University of Rio de Janeiro – Physics Institute*
*marta@if.ufrj.br*
**M. S. O. Massunaga**
*UENF – State University of North Rio de Janeiro*
*shoey@uenf.br*
**G. Rubini**
*UFRJ – Federal University of Rio de Janeiro*
*gustavorubini@if.ufrj.br*

**Abstract**
The entrance to most of the public universities in Brazil depends on the scores obtained by students in Enem, Brazilian National Assessment of High School Education. This exam consists of a written composition and four multiple choice tests (Language, Mathematics, Natural Sciences and Human Sciences) scored by Item Response Theory. It is a high stake exam for students, allowing the use of its results of this exam to provide conditions for monitoring high school learning and quality of education in Brazil. We present a study of the Physics items of Natural Sciences test of Enem over the years 2009 to 2014. The questions were classified based on qualitative categories, providing a profile of the Natural Sciences test: a disciplinary exam (items of Physics, Chemistry and Biology), with mostly qualitative questions, long and time consuming items, only a few of them demanding more complex reasoning and assessing some problem solving skills. The answers of the students were analyzed, using basic statistical exploration and Item Response Theory to draw the item characteristic curve of the Physics items. A comparison was made between the qualitative analysis of the questions and the quantitative analysis for the Physics questions. The study of student performance reveals the low attaining levels of Physics learning at the end of high school, with a small percentage of correct answers on items; students perform worse in questions that require some type of disciplinary knowledge or use of basic mathematical reasoning. It is also possible to observe discrepancies on the results of two subsets of groups: female students perform worse than male students in most items, and students that come from federal and private high school perform significantly better that those from state high school (75% of students in high school attend state or regional schools). In conclusion, the combined analysis of student's performance and content of the items reveal some of the characteristics of student's learning in Brazil, at around 18 years, that can impact Physics Education system.

**Keywords**
Physics Education; Large Scale Assessment; Enem - Exame Nacional do Ensino Médio.

# INTRODUCTION

Large scale assessments are usually employed for system assessment [1,2]. In 1998, following the new Brazilian Constitution at the end of military dictatorship, a national exam – Enem (Exame Nacional do Ensino Médio, or National Assessment of High School Exam) – was created by the Federal Agency on Research and Statistics in Education (Inep) in order to provide students at the end of high school with a tool for "*self assessment of skills and competencies acquired at the end of basic education*" [3]; there is no national curriculum, only recommendations from the National Council on Education, and at the end of secondary level, the students shall master "*the science and technology principles that underpin modern production*" [4]. This exam was a non compulsory test, a general one with no explicit disciplinary content assessment, with 63 questions and a written composition.

Following political changes in 2002 due to presidential elections some major changes occurred in public policies. There was a huge expansion of educational opportunities at university level, together with attempts to improve the quality of basic education by promoting in service teacher' programs and to implement a culture of assessment in education, with indexes and objectives for each school level in the whole country [5].

The expansion of the federal university system that began in 2007 required a change in Enem: its scores were used as part of the selection process for entrance in public universities (all tuition-free) or to obtain full or partial scholarship and loans in private universities. A redesign in the objectives and methodological framework of Enem soon followed: in 2009, the assessment was composed of four tests in general areas (Natural Sciences, Human Sciences, Language and Mathematics) using the three-parameter logistic (3PL) model of Item Response Theory – IRT [6] based in a Reference Matrix comprising five cognitive domains and, for each area, competencies, skills and a list of disciplinary contents. The questions, or items, were prepared by high school teachers and university faculty, composing a huge bank of multiple choice items [7,8].

Enem constitutes, from 2009 to now, a high stake assessment for students in Brazil, with some characteristics that are different from other large scale assessments. Even though it is neither universal nor statistically sampled, it provides a unique tool for research on the diagnostic of Physics (and other disciplines) learning in Brazilian High School allowing to go beyond the general statement on the low quality of science education in the country as revealed by international comparative Pisa results [9]. The huge population of candidates and the number of questions in the exam allow inferences on what students really learned or the real curriculum developed at schools.

The study of this exam poses theoretical questions on the framework to be used. Assessments are made to establish some criteria on the construct to be analyzed – the construct of Enem is revealed in its design and its items that are supposed to measure the latent trait proposed. To gather information on the quality of education, some epistemological choices are usually required. In general, quantitative methods are related to an epistemology of defining a measurement, or some kind of standard based criteria. Qualitative methods, on the other side, are based on description and interpretation of perceptions on the theme by the subjects. So, it is a challenge to combine two ways of thinking, "quality as a measure" and "quality as experience" [10] and use the available data, on the answers given by students, on the questions of the exam and on the socio-economic and personal questionnaires answered by students, to refine the diagnosis on what students have learned about Physics at the end of High School.

This paper describes some of the answers to the question, what can we infer about learning of Physics, using available data from Enem.

**WHAT IS THE EXAM, AND WHAT CAN WE INFER FROM ITS RESULTS**

In the period 2007-2014 there was a huge change in the federal university system in Brazil. The 1988 Brazilian Constitution attributes the responsibility for undergraduate and graduate studies to the federal government, and K-12 studies to state and city governments. The federal universities are public and tuition-free and undergraduate courses take from 4 to 6 years to completion.

The number of federal universities was 45 in 2002 and 63 in 2014, a 40% increase. Most of the new universities are in poorer and interior regions in the country. The number of enrollments in public federal universities undergraduate courses went from 113 thousands to 246 thousands, an increase of about 120%, and in graduate courses the increase was of the order of 300% [5].

The expansion of the federal higher education system and its decentralization brought a series of challenges. One of them was the student´s selection process. In 2009, the Ministry of Education created a national system, in which the students nationally enroll in the federal universities. The selection of students is based on the Enem score. Also, public financing of scholarships and loans to private universities was also transformed in a national process, based on the same scores. That chain of events transformed the National Exam in a very important test for students that want to enter universities.

The exam itself underwent a series of changes in design and methodology. The Natural Sciences exam now comprises 45 multiple choice items and the score is obtained by using a 3PL IRT model.

The exams were prepared following a complex Reference Matrix [7] composed of five cognitive domains (domain over languages, understanding phenomena, confronting problem-situations, argument, elaborating proposals), seven competences, 30 skills, and knowledge objects, or disciplinary contents. Each item must assess a single skill [8].

In Natural Sciences, the competences are
1. To understand natural sciences and technologies associated as human constructions, perceiving their roles in production processes and in social and economic development of mankind.
2. To identify the presence and apply the technologies associated with natural sciences in different contexts.
3. To associate interventions that result in environmental degradation or conservation to productive and social processes and to scientific-technological instruments or actions.
4. To understand the interactions between organisms and environment, particularly those related to human health, connecting scientific knowledge, cultural aspects and individual characteristics.
5. To understand methods and procedures pertinent to natural sciences, and apply them in different contexts.
6. To appropriate of Physics knowledge to, in problem-situations, interpret, evaluate or plan scientific-technological interventions.
7. To appropriate of chemistry knowledge to, in problem-situations, interpret, evaluate or plan scientific-technological interventions.
8. To appropriate of biology knowledge to, in problem-situations, interpret, evaluate or plan scientific-technological interventions.

The disciplinary contents in Physics are composed of
1. Basic and fundamental concepts.
2. Movement, equilibrium and the discovery of physical laws.
3. Energy, work and power.
4. Mechanics and how the universe works.
5. Electromagnetic phenomena.
6. Oscillations, waves, optics and radiation.
7. Heat and thermal phenomena.

The learning definition, or construct, of Enem is subject to debate in the country; there is a huge pressure in educational areas in order to change the Reference Matrix. The matrix poses some problems on unidimensionality requirement on items as postulated by Item Response Theory. The list of skills and competences are somehow superposed, and it is difficult to classify unambiguously the skills and competences assessed by the items.

Nonetheless the exam in its new form has now completed 8 editions and microdata are available for 6 of them. The information that can be gathered from its results is very relevant to understand the situation of learning by young people in the country.

**HANDLING ENEM DATA AND ITEMS**

The exam takes place at the same two days and at the same time in the entire country, and consists of 45 questions for each area in four versions, with the same questions presented in different numbering order. The applications are made online, about four months before the weekend in which the exam is taken.

The federal agency Inep makes available, usually from two to three years after the exam, all the data collected from students: answers to all the items on the exams, and to the questionnaires about personal information and school information of the applicants. These data can be downloaded and analyzed. Also, the exams are available on line.

The microdata is gathered and organized so that exploratory statistical analysis could be performed with statistical packages, namely SPSS [11].

Some filters were applied to these data.

*Participants* are the ones that, after application, take the first two exams, in the first day (Natural and Social Sciences exam).

In the questionnaire, students shall declare whether they are graduating in High School in the year of the exam; using this declaration, it is possible to choose the group classified as "students concluding High School" at the year of the exam. The data were also filtered so that students that did not respond the other exams or that did not present a valid written composition were excluded from the sample. This group is the subset used in the analysis, as representative of the universe of the students graduating in High School.

The microdata were organized so that all the questions are referred to one of the four versions of the Natural Sciences exam, characterized by color – the numbers are referred to the ones in the blue exam.

In Table 1 the numbers involved in the quantitative data are shown. The applications to the exam are of the order, in 2014, of 8.7 million students. The participants are from 2.3 to 5.6 million from 2009 to 2014, around 65% of the applications. The group studied, participants concluding High School with all exams valid, has from 0.8 to 1.4 million in number, circa 30% of the total of participants, which is an indicative of the huge deficit in university enrollments in Brazil.

From official Brazilian data [12] the number of students graduating in high school every year from 2009 to 2012 is almost constant, of the order of 1.8 million. Table 1 unfolds a crescent participation of this population on the exam signaling a rise in its relevance to the students.

**Table 1.** Number of students participating in Enem from 2009 to 2014: applications, participants and the ones that were concluding High School at the year of the exam.

|  | Enem 2009 | Enem 2010 | Enem 2011 | Enem 2012 | Enem 2013 | Enem 2014 |
|---|---|---|---|---|---|---|
| Applications | 4,148,721 | 4,626,094 | 5,380,856 | 5,791,065 | 7,173,563 | 8,722,248 |
| Participants | 2,330,534 | 3,101,455 | 3,670,241 | 3,942,639 | 4,908,306 | 5,633,954 |
| Participants concluding High School | 864,827 | 1,059,227 | 1,174,429 | 1,205,063 | 1,326,681 | 1,374,821 |

*(Source: the authors, with data provided by Inep)*

The available data includes scores – estimated by 3PL IRT model – which are used to allow for decision on entering public universities and obtaining scholarships or financing to private universities.

In this work, the 3PL model was fitted in R [13] using the "mirt" package [14]. The item parameters, which are not provided by Inep, were fitted by the marginal maximum likelihood estimation using the EM algorithm [15]. Students' abilities were estimated by the expected a posteriori method, which is also used by Inep to calculate the Enem scores [16]. This fit provided the estimated item parameters used to reproduce the Item Characteristic Curve (ICC) or Function presented here, and also the Test Information Curve [6].

The official Inep scores of the candidates were classified in ten equally sized groups. In each of these groups the fraction of correct answer per item is obtained, so that an empirical ICC can be drawn for comparison between data and model.

To make inferences based on quantitative results, it is necessary to discuss the content of the items [17]. The construct of the exam is composed of competences and skills, disciplinary contents and cognitive domains. The items on the exam shall be analyzed in order not only to provide inferences on what is being learned, but also if they measure what they are supposed to measure.

The items or questions of the Natural Sciences tests from 2009 to 2014 were analyzed following categories proposed by Gonçalves and Barroso [18]. The classification was made independently by each researcher and then compared and discussed in order to obtain consensus.

The first category was the disciplinary subject. Although the Natural Sciences exam is proposed as an interdisciplinary test [19] and the curriculum in Brazilian High School is based on an interdisciplinary proposal and legal basis [4], the schools teach Natural Sciences subjects as separated disciplines – Physics, Chemistry and Biology – during the whole three years of high school. The interdisciplinary then is presumably acquired by assuming that item assessment is based on situations of real life presented as questions [20] based on competences and skills desirable at the end of high school. The interdisciplinary concept in science teaching at high school is a controversial theme, as presented in Czerniak and Johnson [21] and Lederman and Niess [22]; even inside Natural Sciences, each of its constituent disciplines possesses conceptual, epistemological and procedural differences. In this sense, it is reasonable to provide a disciplinary classification for each item. The disciplinary category was chosen based on the subject content needed to correctly answer the item.

The Natural Sciences exam presents 45 items, and the disciplinary content is almost equally distributed. In Table 2, the disciplinary distribution is shown.

**Table 2.** Disciplinary categorization of Natural Sciences Exam from Enem 2009 to Enem 2014.

| Year / Discipline | 2009 | 2010 | 2011 | 2012 | 2013 | 2014 | Average per year |
|---|---|---|---|---|---|---|---|
| Physics | 16 | 16 | 15 | 16 | 16 | 15 | 15.7 |
| Chemistry | 9 | 13 | 13 | 15 | 14 | 15 | 13.2 |
| Biology | 20 | 16 | 17 | 14 | 15 | 15 | 16.2 |

*(Source: the authors, with data provided by Inep)*

Other categories used were related to content analysis; items can be qualitative or quantitative (with some arithmetic or algebraic reasoning needed), present information in graphic or imagetic form, be associated with practical aspects of science teaching, require some specific knowledge of science (content knowledge) not present in the text, require higher cognitive skills (as in the case of real problems with connections between concepts) and other possibilities.

For each Physics item, the percentage of choice for each alternative – both the correct one and the distractors – was obtained and the empirical data was compared to the model ICC.

All the information was gathered to provide a deeper profile on Physics learning by Brazilian High School graduating students. Next, some selected results are presented in order to provide the scenario for the conclusions.

**SOME GENERAL RESULTS**

The score on the exam was normalized in the 2009 exam for students concluding high school: mean 500 and standard deviation 100. In Table 3, general data on Natural Sciences exam from 2009 to 2014 (mean score and standard deviation) are shown.

**Table 3.** General results on Natural Sciences Exam from Enem 2009 to Enem 2014 – IRT Scores

|  | Enem 2009 | Enem 2010 | Enem 2011 | Enem 2012 | Enem 2013 | Enem 2014 |
|---|---|---|---|---|---|---|
| Mean | 501.1 | 485.0 | 466.5 | 473.0 | 473.3 | 487.6 |
| Std. Deviation | 98.5 | 81.2 | 85.1 | 79.9 | 75.6 | 74.5 |

*(Source: the authors, with data provided by Inep)*

It is interesting to compare these scores with the number of correct answers on the exam, providing a score based on Classical Test Theory – CTT. There were 45 items in each exam, and the sum of correct answers, normalized to 10, is shown in Table 4. It can be observed that the mean value of correct answers in the exam is only one third of the total number of questions, and that this number is slowly decreasing along the years.

**Table 4.** General results on Natural Sciences Exam from Enem 2009 to Enem 2014 – Grade corresponding to Classical Test Theory – total of correct answers, from 0 to 10

|  | Enem 2009 | Enem 2010 | Enem 2011 | Enem 2012 | Enem 2013 | Enem 2014 |
|---|---|---|---|---|---|---|
| Mean | 3.4 | 3.2 | 3.2 | 3.0 | 2.6 | 2.8 |
| Std. Deviation | 1.2 | 1.3 | 1.3 | 1.3 | 0.9 | 1.1 |
| Minimum | 0.0 | 0.0 | 0.0 | 0.0 | 0.0 | 0.0 |
| Maximum | 10.0 | 9.8 | 9.8 | 9.8 | 9.8 | 10.0 |

*(Source: the authors, with data provided by Inep)*

In Figure 1, the histograms of the scores of IRT and of the CTT grades (sum of the correct answers, in 45 questions, normalized from 0 to 10) are show for the 2013 Natural Sciences exam. These figures show that the number of correct answers is low. The correlation between these two scores is shown in Figure 2 – $R^2 = 0.762$ – and this graph shows an expected result as the 3PL model scores are not dependent on the number of correct answers but on the student's pattern of responses. For example, students A and B may have the same number of correct answers but their 3PL model scores might be different based on which questions each one has responded correctly.

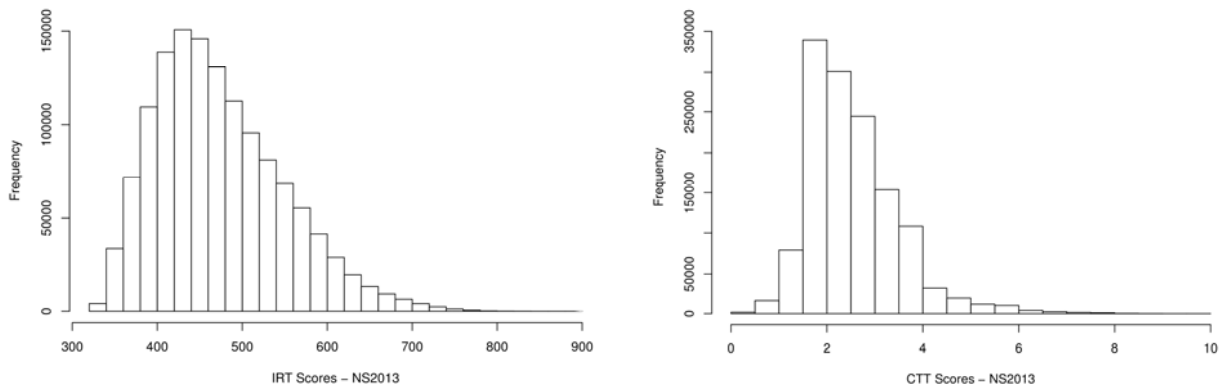

**Figure 1.** Histograms of student's scores on Natural Sciences test of 2013; from left to right, the IRT score histogram and the CTT score histogram (observed, normalized from 0 to 10).

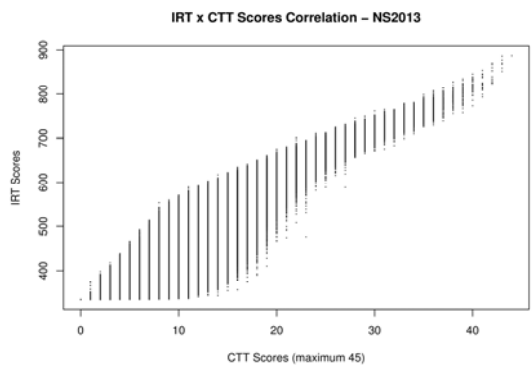

**Figure 2.** Correlation between IRT scores and number of correct answers (observed non-normalized CTT score) on Natural Sciences test of 2013; $R^2=0.762$.

In Figure 3, the correlation between the official IRT scores presented in the microdata and the scores calculated using the R-mirt package that provided the item parameters is shown, for the Enem 2009 data. It can be shown that the two models are correlated with $R^2=0.9992$. This allows us to infer that, even though the official item parameters are not publicly available, our estimated parameters are reasonably valid and therefore it is possible to extract useful information from them.

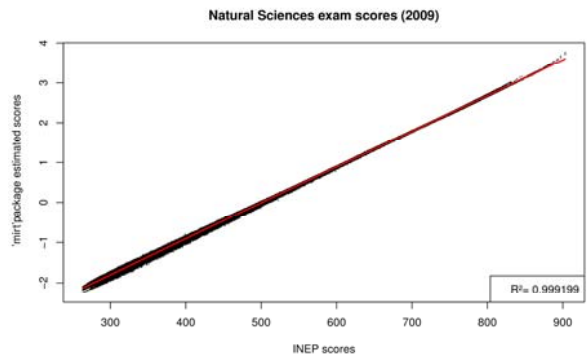

**Figure 3.** Correlation between IRT obtained by the federal agency Inep and using R-mirt package.

In Figure 4, the Test Information and Standard Errors on the Natural Sciences exam of Enem 2014 are presented. The information curves from Enem 2009 to 2013 are similar to this one; it can be seen that the scores, in statistical units, are precise in the region of about 1.5 to 2.5 and that the exam is adequate for the selection of students for public universities, but it is not an accurate measure for regions below 0.

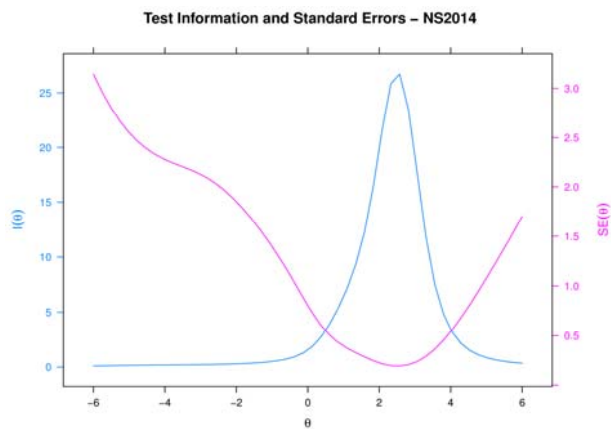

**Figure 4.** Test information curve (in blue) and standard errors in scores (in pink) of the Natural Sciences exam of Enem 2014.

These general characteristics of the Natural Sciences exam reveal a surprising low number of correct answers and an adequate information for university selection.

**SOME DISCREPANCIES ON SOCIAL GROUPS REVEALED BY THE EXAM**

Gender differences can be studied. Females are well represented in the exam, being the majority of the candidates as shown in Table 5. However, the female group has an average score lower than the male group as shown in Table 6.

**Table 5.** Percentage of female and male candidates concluding high school per year of exam.

|        | Enem 2009 | Enem 2010 | Enem 2011 | Enem 2012 | Enem 2013 | Enem 2014 |
|--------|-----------|-----------|-----------|-----------|-----------|-----------|
| Female | 61.3      | 60.9      | 60.1      | 59.6      | 59.0      | 58.6      |
| Male   | 38.7      | 39.1      | 39.9      | 40.4      | 41.0      | 41.4      |

*(Source: the authors, with data provided by Inep)*

**Table 6.** Female and male IRT group scores (mean ± standard deviation).

|        | Enem 2009     | Enem 2010     | Enem 2011     | Enem 2012     | Enem 2013     | Enem 2014     |
|--------|---------------|---------------|---------------|---------------|---------------|---------------|
| Female | 488.2 ± 94.6  | 475.9 ± 76.3  | 456.7 ± 80.8  | 462.4 ± 74.9  | 464.8 ± 72.3  | 479.3 ± 70.4  |
| Male   | 521.5 ± 100.9 | 499.0 ± 86.4  | 481.3 ± 89.2  | 488.7 ± 84.3  | 485.4 ± 78.4  | 499.4 ± 78.4  |
| Total  | 501.1 ± 98.5  | 485.0 ± 81.2  | 466.5 ± 85.1  | 473.0 ± 79.9  | 473.3 ± 75.6  | 487.6 ± 74.5  |

In Figure 5, the amount of female and male scores per decil (the group was divided into 10 equal groups, the 10% of lower grades, classified as 1, to 10% to higher grades, classified as 10) in 2013 and the female percentage per decil score from 2009 to 2013 are shown. These graphs, more than showing that the groups score distribution of abilities are distinct, exhibit differences in the male/female proportion by decil group, with relatively more female presence in lower scores and male presence in higher scores.

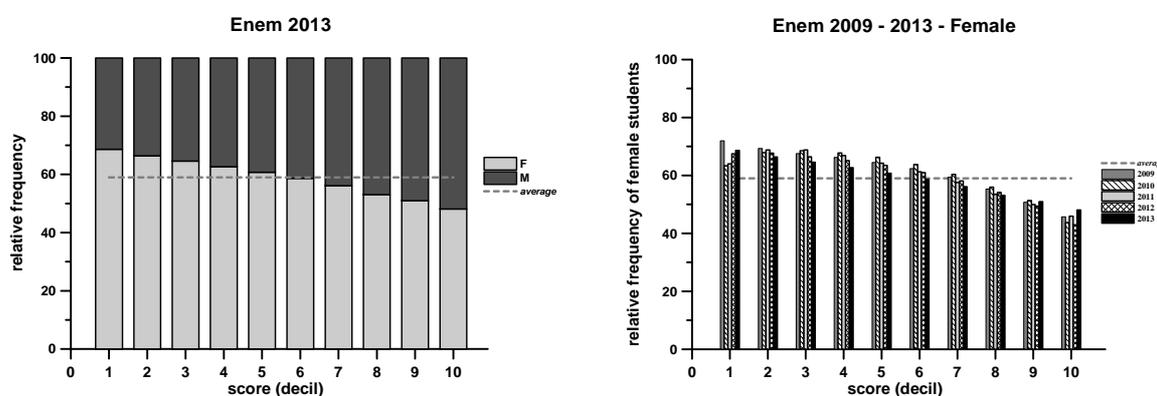

**Figure 5.** (left) For each decil group on scores, Enem 2013, the percentage of female (light grey) and male (dark grey) students, with a line showing the average percentage of female in the universe studied. (right) Percentage of female students in the exam from Enem 2009 to Enem 2013.

Brazilian high schools are mainly public ones, with financing from federal, state or city governments; according to Brazilian laws, the responsibility for secondary upper schools is due to the states of the federation. There also private schools.

In Table 7, the distribution of students in the various types of schools in the country is presented. It can be observed that three quarters of the students concluding high school come from state schools. In Table 8, score averages are shown for each type of school. State schools have the lowest average performance.

**Table 7.** Percentage of students per type of high school (public: federal, state or city, and private) concluding high school per year of exam.

| | Frequency (%) by school type | | | | | |
|---|---|---|---|---|---|---|
| School  | 2009 | 2010 | 2011 | 2012 | 2013 | 2014 |
| Federal | 1.8  | 1.8  | 1.9  | 2.0  | 2.0  | 2.2  |
| State   | 74.5 | 74.6 | 75.1 | 73.9 | 74.9 | 72.5 |
| City    | 1.6  | 1.4  | 1.2  | 1.1  | 1.0  | 1.0  |
| Private | 22.1 | 22.3 | 21.9 | 23.0 | 22.1 | 24.4 |
| Total   | 100  | 100  | 100  | 100  | 100  | 100  |

**Table 8.** Type of high school IRT group scores (mean ± standard deviation).

| | Score mean and s.d. by school type | | | | | |
|---|---|---|---|---|---|---|
| School | 2009 | 2010 | 2011 | 2012 | 2013 | 2014 |
| Federal | 600.5 ± 92.4 | 566.3 ± 77.5 | 548.7 ± 83.5 | 553.9 ± 81.5 | 542.8 ± 81.9 | 559.0 ± 76.5 |
| State | 475.6 ± 83.4 | 464.0 ± 69.5 | 444.5 ± 71.4 | 450.8 ± 64.5 | 453.2 ± 60.1 | 469.6 ± 61.4 |
| City | 483.2 ± 86.8 | 475.1 ± 72.8 | 460.5 ± 75.2 | 464.5 ± 70.5 | 464.6 ± 66.5 | 483.4 ± 66.2 |
| Private | 580.4 ± 99.4 | 549.4 ± 80.7 | 535.4 ± 88.2 | 537.7 ± 84.9 | 535.3 ± 84.6 | 535.0 ± 84.7 |
| Total | 501.1 ± 98.5 | 485.0 ± 81.2 | 466.5 ± 85.1 | 473.0 ± 79.9 | 473.3 ± 75.6 | 487.7 ± 74.5 |

Enem candidates come mainly from public state schools (~ 75% of students) followed by private schools (~21%); public municipal and public federal schools candidates are the minority, each one with less than 2% of candidates (Table 7). Table 8 shows clearly that federal schools have best results overall, followed by private schools. Despite the low number of candidates, public federal schools are important because, generally, they have characteristics that differentiate them from the other public schools; for example, their teachers, compared to public state school ones, have higher salaries (equivalent to university salaries) and their career encourages the pursuit for graduate studies. And their students perform significantly better.

From official data [12], circa 85% of high school students are enrolled in state schools (versus ~75% in Enem) and 13% in private schools (versus almost one fourth in Enem), so it is evident that a significant percentage of state school students do not even try to enter university. We can infer that there is an educational exclusion mechanism that begins even before the Enem signing up process.

Those discrepancies in scores (gender and school type) must be further analyzed in order to investigate if the Enem exam is fair and they are simply the consequence of different distribution of abilities (impact) of if some items exhibit differential item functioning [23] which could bias the test results.

Some other differences can be observed (e.g., related to regions in the country) but these are the two considered as with utter relevance; Pisa 2006 results do not showed gender differences (at beginning of high school), but Enem consistently shows these differences (in test and item performance, in all the years considered). Also, the inequality of Brazilian education related to the type of school and region of the country is shown to be reflected in Enem results.

**PHYSICS ITEM CHARACTERISTIC CURVES**

In order to make inferences about Physics learning at the end of high school, the items shall be analyzed one by one. The content of the items and what they are measuring are essential to understand the numbers on the performance of students [17].

The analysis took into account two quantitative information from IRT and CTT: the item characteristic curves and distractor analysis.

In Figure 6, the ICC of all the items in Natural Sciences Enem 2010 are shown. Curves marked with an asterisk symbol (*) correspond to the categorized Physics items. These curves, obtained in mirt R package, show the probability of choosing the correct answer versus students' ability in statistical units.

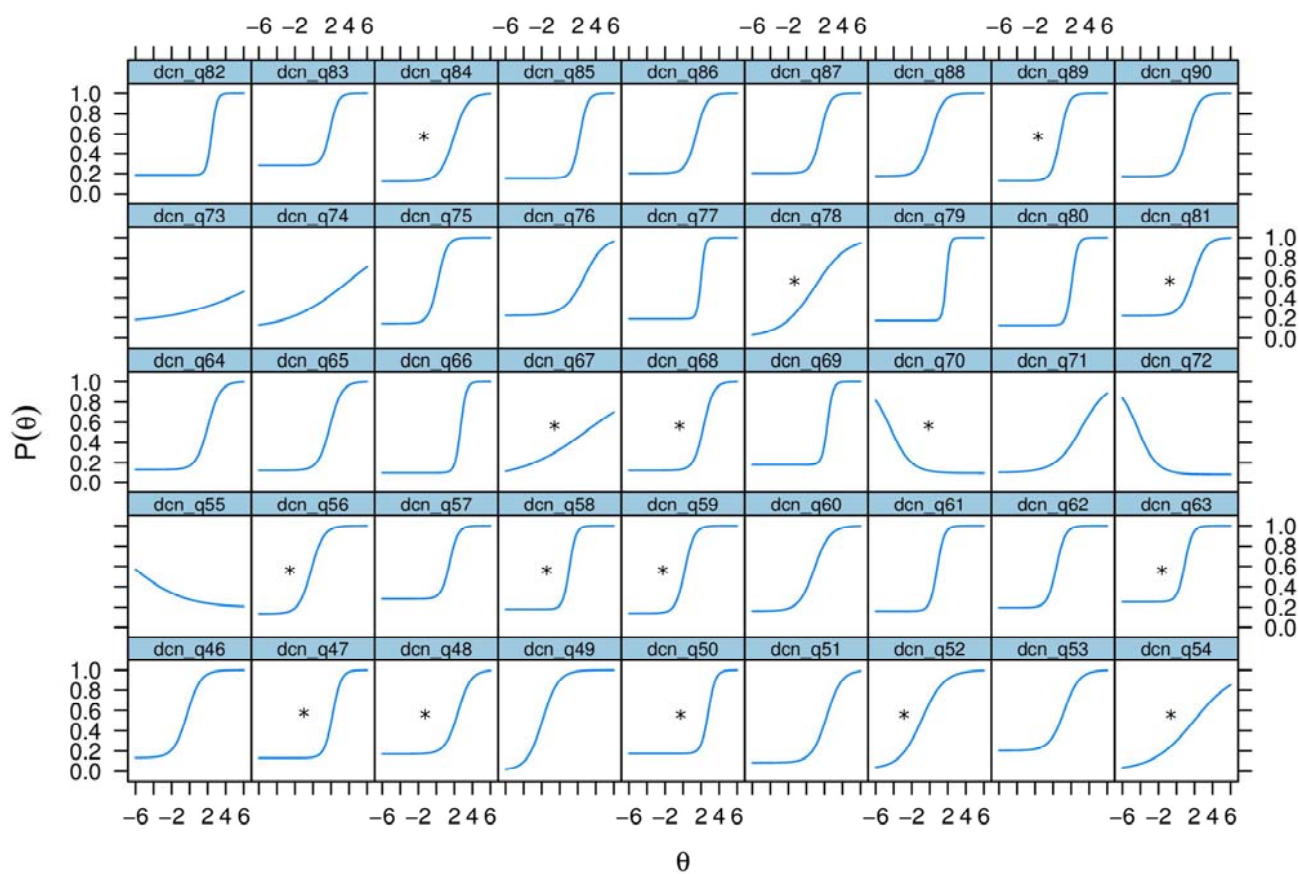

**Figure 6.** Item characteristic curves for 2010 Natural Sciences Exam. Physics items are marked.

It can be observed that some items do not have the expected behavior according to the 3PL IRT model. For example, items 55 and 70 do not monotonically increase as postulated by IRT; similar behaviors in some items were observed in the analysis of other years.

These observations can lead to the need of a thorough reexamination of the whole exam, so that the interpretation of these curves as a possible violation of the unidimensionality postulate of IRT can be ruled out. Only a combined study of the content of the item and performance of students can give a reasonable explanation on exactly what is happening with the psychometric properties of the items.

**EXAMPLE OF ANALYSIS – PHYSICS ITEMS**

All the Physics items were analyzed both in their content (the Physics subject, the structure of the text, the conceptions presented in the alternatives, the usual mistakes, and so on) and in the performance of students (percentage of students choosing each alternative, the total number and distribution of correct answers, the ICC of the item and so on), allowing us, by comparing content and performance, to make some inferences on what the exam reveals about Physics learning in high school.

In order to illustrate the reasoning used in our inferences about Physics learning in high school, two Physics items were selected.

In Figure 7, item 47 of Enem 2012 is presented. This item assesses the relation between force, pressure and area in a real situation and is composed of three parts: the situation, the command (direct question) and 5 alternatives (one considered to be the correct one, A, and four distractors). In Table 9, the percentage of choice of each one of the answers is shown. Figure 8 shows ICC curves: the line shows the model curve (mirt parameters) and the empirical data (obtained by dividing students in ten equal numbered groups of scores and calculating the fraction of correct answers in each group); the scores are presented in statistical units (difference, in standard deviation units, of the mean value 0) and probability of having the correct answers is given as the fraction of total number.

> An environmental problem experienced by agriculture today is soil compaction due to intense traffic of heavy machinery reducing crop productivity.
>
> One form of preventing soil compaction is replacing the tractor tires with tires that are
>
> **(A)** wider, reducing the pressure on the ground.
> (B) narrower, reducing the pressure on the ground.
> (C) wider, increasing the pressure on the ground.
> (D) narrower, increasing the pressure on the ground.
> (E) taller, reducing the pressure on the ground.

**Figure 7.** Item number 47 of Natural Sciences exam of Enem 2012 (English version by the authors).

**Table 9.** Percentage of choice of each one of the alternatives of Question 47 – CN 2012

| NS2012 - Q47 | |
|---|---|
| option | % |
| **A** | 43.9 |
| B | 32.6 |
| C | 3.8 |
| D | 2.6 |
| E | 16.9 |
| null / NA | 0.2 |

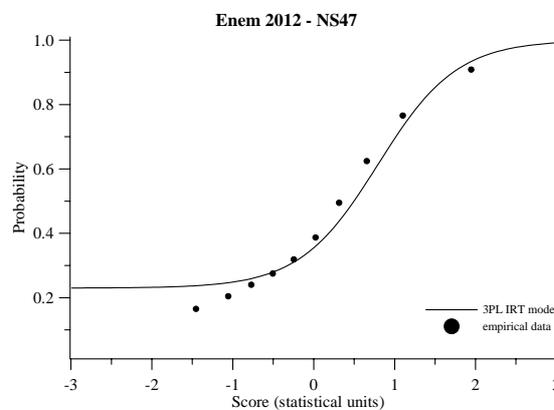

**Figure 8.** Item 47 Characteristic Curve of 2012 Natural Sciences test – empirical data and model.

This item, NS2012-Q47, assesses the relation between pressure, force and area. Circa 44% of the students choose the correct answer (A), connecting an increase in the area to lowering the pressure exerted by a constant force. The other two options that mention a decrease in pressure are also attractive to students: option (B) by 33% and (E) by 17%. So a large amount of students know that the problem shall be solved by reducing the pressure on the tires (92%), but less than half of them seem to have a clear conceptual explanation on this knowledge.

The ICC shows that the item is difficult (item difficulty 0.8) and its discrimination is a good one (discrimination 2.1). Also, pseudo-chance is very high, 0.23. It is seen that empirical data is reasonably fitted by the 3PL model employed in the description.

In Figure 9, another item (number 70) of Enem 2011 is presented. This item proposes the discussion of how to make a lamp shine with a battery, electrical wires and a bulb. It is also composed of three parts: the situation, the command (direct question) and 5 alternatives (one considered to be the correct one, D, and four distractors). In Table 10, the percentage of choice of each one of the answers is shown. Figure 10 shows ICC curves: the line shows the model curve (mirt-irt parameters) and the dots represent the empirical data.

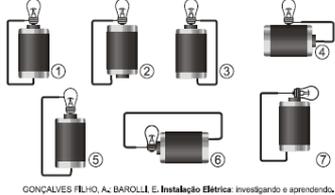

A curious student, thrilled by the electrical circuit lesson at school, decides to disassemble his flashlight. Using the lamp and the batteries, taken from the equipment, and a wire with stripped ends, he makes the following connections intending to light the lamp:

Based on the schemes shown, in which of the situations does the lamp light?

(A) (1), (3), (6)
(B) (3), (4), (5)
(C) (1), (3), (5)
**(D)** (1), (3), (7)
(E) (1), (2), (5)

**Figure 9.** Item number 70 of Natural Sciences exam of Enem 2011 (English version by the authors).

**Table 10.** Percentage of choice of each one of the alternatives of Question 70 – NS 2011

| NS2011 - Q70 | |
|---|---|
| option | % |
| A | 25.2 |
| B | 6.1 |
| C | 27.7 |
| **D** | 34.7 |
| E | 6.0 |
| null / NA | 0.4 |

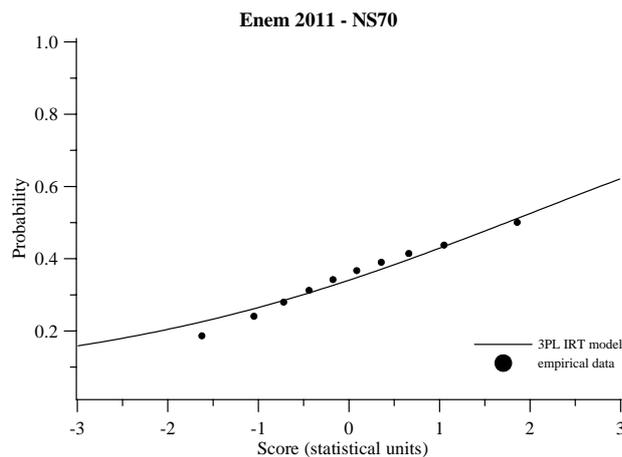

**Figure 10.** Item 70 Characteristic Curve of 2011 Natural Sciences test – empirical data and model.

This item, NS2011-Q70, assesses some practical skills on simple electric circuits. Circa 35% of the students choose the correct answer (D), which presents 3 possible connections, 1-3-7, between positive and negative terminals of the battery and the lamp bulb. At the same time, the scheme 5 – connecting one of the terminals with the glass of the bulb – presented in options (B), (C) and (E) is chosen by 40% of students, showing the lack of familiarity with electric conducting properties of materials. Choices (A), (C) and the correct one (D) are different only in the last scheme – 6, 5 or 7.

Apparently some kind of "multiple choice answering technique" was used, eliminating options (B) and (E), with an almost random choice for the three similar options.

The ICC curves show that this item has low discrimination capacity (discrimination 0.41) and it is not easy to visualize the 2.0 difficulty parameter. Also, pseudo-chance parameter is low (0.05).

This theme reveals one of the known characteristics of Physics teaching in Brazil: the almost total absence of practical activities at school. And the results on the item reveal students' difficulties on simple technical details, showing a lack of practical activities and of school laboratory experience: there is a high percentage of candidates that connected the wire to the bulb.

The combined analysis presented in these two examples was developed on all Physics items from 2009 to 2014, and part of is presented elsewhere [18, 24, 25].

**FINAL COMMENTS**

The results shown here are part of a more complete analysis of the Enem exams. From the analysis, some conclusions can be presented.

Enem is not a typical large scale assessment: it is not universal, nor uses any sampling technique, but is a high stake assessment for Brazilian students – it provides conditions to entering the university system in the country. The number of applications and participation on the exam is large fraction of the total number of students concluding high school.

The exam has a multidimensional Reference Matrix, with a construct of Science Learning not completely interdisciplinary as proposed in its documentation and in legal basis. This might conflict with the unidimensionality assumption of the IRT model used in the analysis.

The exams are prepared at the federal agency responsible for the assessment; they indirectly reflect what university faculty and high school teachers expect on the Physics learning at high school level, which is not completely aligned with what is proposed on legal curricular orientations.

The results attained by students show an unexpected pattern of performance: the mean number of correct answers is of the order of one third of the total of items. The classification based on item content provides some inferences on learning.

Items that require

- graphical interpretations show that competences related to non verbal languages skills are not attained at the end of high school;

- some quantitative reasoning have a percentage of correct answers well below average;

- problem solving skills and multiple cognitive steps have less correct answers than what would be attained by entirely random choice by students;

- specific Physics content knowledge rather than a mere interpretation of written information have the percentage of correct answer below global average;

- some laboratory practice reveal that practical activities seem to be absent from teaching.

These are some of the inferences that could be obtained in this analysis so far. It is possible to hypothesize a gap between the construct of the exam and students learning.

Significant difference in performance by gender and type of school are observed.

And students' performance shows the inequity and inequality in the country educational system.

**ACKNOWLEDGMENTS**

The authors are grateful for the partial support of MEC/CAPES through Grant No. 76/Observatório da Educação.